\documentclass[aps,pra,preprint]{revtex4-1}
\usepackage{graphicx}
\begin{document}
\providecommand{\boldsymbol}[1]{\mbox{\boldmath $#1$}}
\providecommand{\tabularnewline}{\\}
\providecommand{\He}[2]{$^{#1}$He$_{#2}$}
\title{Benchmark helium dimer and trimer calculations with a public few-body code.}
\author{Vladimir Roudnev and Michael Cavagnero}
\begin{abstract}
We present detailed calculations of bound and scattering states of dimers and trimers of He to produce highly accurate data
and to test a non-relativistic three-body code currently in development for public distribution.
For these systems, uncertainties and inaccuracies in the fundamental constants frequently used in published works can substantially exceed numerical errors. Our benchmark calculations include specific estimates of the numerical accuracy of the calculations, and also explore sensitivity to fundamental constants and their uncertainties. The use of an inexact coupling constant in the previous calculations leads to 0.08\% error for the ground state energy, 0.3\% error for the excited state energy and up to 0.15\% error for the atom-dimer scattering length in the system of three $^4$He atoms. The corresponding errors for the unsymmetric \He{4}{2}\He{3}{} system are 0.3\% for the bound state energy and 0.03\% for the atom-dimer scattering length.
\end{abstract}
\maketitle
\section{Introduction}

Experimental investigations of He dimers and trimers \cite{HeExp1,HeExp2,HeExp3} in
the 1990's generated considerable theoretical interest in these weakly-bound few-body systems.
Small He clusters are now considered one of the most extensively studied molecules --
theoretically and computationally \cite{Az79,HFD-B,LiuMcLean,LM2M2,TTY,VanMourik,SAPT2007,Pandharipande,Glockle,ELG,NFJ,MSK97,KMS98,KM99,RoudnevYakovlev,
BG,HeNeAr,Filikhin,RoudnevHeScat,Braaten,Penkov,RoudnevFBS,Carbonell,PlatterPhillips,VK,Yarevsky,SunoEsry2008,KolganovaMotovilov}
(see also an extensive review \cite{KolganovaMotovilov}).
Many semi-empirical and ab initio potential models have been proposed for the interaction of He atoms\cite{Az79,HFD-B,LiuMcLean,LM2M2,TTY,VanMourik,SAPT2007}.
These potential models have been used as input in a number of two- three- and even four- body quantum calculations\cite{Filikhin,Carbonell}.
The numerical results of these investigations are, for the most part, in agreement, with small remaining
discrepancies due simply to limits of numerical accuracy. Nevertheless, we wish to revisit, reproduce
and check these previous results. While small clusters of He are structurally quite simple, they are also delicate, with very weak
binding, and so are challenging to model numerically.
Given the extensive set of known potential models and the number of results published, it seems natural to use bound and scattering
states of two and three He atoms to benchmark the accuracy of existing and future few-body quantum-chemical software.

The near-threshold state of the He dimer is extremely sensitive to the details of the interaction potential.
Visually indistinguishable potentials can lead to qualitatively different expectations for the properties of the He dimer.
The three-body states are likewise expected to be very sensitive to the details of the interaction, since their binding
is similar to that of Efimov states. One immediate result of our investigations is that, even for a fixed potential,
He cluster observables can vary substantially depending on the accuracy of the fundamental constants used in the calculations.
Unfortunately, the fundamental constants and conversion factors published by the developers of potential models and early
He cluster calculations -- and, therefore, widely used in subsequent publications -- have intrinsic systematic errors
that prevent them from being used for benchmark purposes.

In this paper, we consider three questions: To what extent is the accuracy of near-threshold calculations limited by
knowledge of the fundamental constants? To what extent can the known discrepancies in the published results be explained by possible
use of inexact coupling constants? Can our three-body calculations reach the ``natural'' accuracy limit set
by the uncertainties of the known fundamental constants?
Answers to these questions will allow us to benchmark a quantum few-body code being prepared for public release, with intended applications to a wider variety of atomic and molecular few-body systems.

The detailed technical description of the three-body code we use for the calculations will be made elsewhere;
generally, the techniques used in these calculations are similar to the ones employed in
\cite{RoudnevFBS,RoudnevHeScat,RoudnevYakovlev}.
In the current stage of development the code is not yet completely ready for an extensive public distribution. Interested readers, however, are encouraged to contact the authors for obtaining the current development version.

We assume throughout this paper that the system of He atoms can be described by a single-channel pairwise interaction. We, therefore ignore the contribution of three-body forces. This contribution can, substantially exceed the numerical accuracy of the calculations presented. Our goal, however, is to provide an exemplary data set based on known potential models rather than to perform the most realistic calculation of the physical system. We also ignore all the off-diagonal non-adiabatic corrections not included in the single-channel potential model.

In the following sections we shall give a short description of the problem, analyze the uncertainties that enter the benchmarking problem, give a review of published results and provide our results, which account for all known physical uncertainties.

\section{Physical uncertainties in two-body observables}

Consider the Schr\"odinger equation for a two-body system
\[
(-\frac{\hbar^{2}}{2m}\nabla^2+cV(r)-E)\Psi=0\ \ \ .
\]
Here, $c$ is a conversion factor between the energy units of the
potential (provided by the various developers of two-body potentials) and our chosen ``natural'' units, as specified below. This conversion
factor, specified with varying degree of precision by different authors, is usually a simple function of fundamental constants. Rescaling
the energy we reduce the equation to
\[
(-\nabla^2+v(r)-z)\Psi=0\ \ \ ,
\]
where $v(r)=c\frac{2m}{\hbar^{2}}V(r)$ and $z=\frac{2m}{\hbar^{2}}E$.
To analyze the sensitivity of the state $\Psi$ to the inaccuracy
of the fundamental constants and other conversion factors, we introduce a (dimensionless) coupling constant
$\lambda$ and treat the eigenvalue $z$ as a function of
the coupling constant
\[
(-\nabla^2+\lambda v(r)-z(\lambda))\Psi=0\ \ \ .
\]
Knowing the derivative of the energy eigenvalue $\frac{d}{d\lambda}z(\lambda)\vert_{\lambda=1}$
one can estimate the sensitivity of the dimer bound state, or scattering phase shifts, to
the degree of precision of the fundamental constants of interest.

Many model potentials of rare gas atom interactions were published in units of temperature.
The conversion factors between atomic units of energy (typically employed in calculations)
and temperature units are, however, not always provided. In some cases, when the corresponding conversion factors are published,
the values suggested by the authors differ significantly from the values recommended by NIST. For instance,
the recommended value for the TTY potential is 3.1669$\times$10$^{-6}$~a.u.~K$^{-1}$, which differs from
the recommended value of 3.1668154$\times$10$^{-6}$~a.u.~K$^{-1}$ in the 5th significant figure. Even such small
differences can appreciably affect the accuracy of derived results owing to the weakly-bound nature of few-body
clusters, and can make benchmarking and comparison of alternative computational strategies difficult.

Many three-body calculations of He clusters \cite{Glockle,KolganovaMotovilov,RoudnevYakovlev,Carbonell}
have been performed, assuming \AA\  for distance and the conversion coefficient
$\frac{\hbar^{2}}{2\mu k_{b}}\approx12.12$~\AA$^{2}$K.
Keeping only 4 significant figures may lead, however, to significant inaccuracy in 2-body observables.
Let us estimate the accuracy of this coupling constant based on the best available data
from the NIST database. Based on the following recommended values
\[
2\mu=m_{^{4}He}=(4.0026032497\pm0.0000000010)\ a.m.u.
\]
\[
\hbar=(1.054571628\pm0.000000053)\times10^{-34}Js\]
\[
k_{B}=(1.3806504\pm0.0000024)\times10^{-23}JK^{-1}
\]
\[
1 a.m.u.=(1.660538782\pm0.000000083)\times10^{-27}kg
\]
the value of the conversion factor for two $^{4}$He atoms should be taken as
$\frac{\hbar^{2}}{2\mu k_{b}}=(12.11928\pm0.00002)\ K$\AA$^{2}$.
The major source of the uncertainty in this conversion factor is the Boltzmann constant, followed
by the Planck constant and the mass unit.

The results of our calculations of the binding energy
and effective range expansion parameters for two He atoms are summarized in Table~\ref{tab:TwoBodyLowEnergyParameters}.
For conversion of energy between K and atomic units we use the value of the
Boltzmann constant $k_B= 3.1668154\times 10^{-6}\ \ a.u./K$
We provide two types of results. The first group of calculations is
performed with the effective coupling constant used in previous three-body
calculations, the second type is done with the best values recommended
by NIST. The calculations are done in atomic units. For conversion between K and a.u. (the factor "c" discussed above) we also
use the values recommended by NIST. This explains the difference between the values reported here
and previously published results for the TTY potential.
\begin{table}[!ht]
\begin{tabular}{ccccccc}
\hline
 & \multicolumn{3}{l}{$\frac{\hbar^{2}}{m}\equiv12.12\ $K\AA$^{2}$} & \multicolumn{3}{l}{$\frac{\hbar^{2}}{m}\equiv12.11928\
$K\AA$^{2}$}\tabularnewline
\hline
Potential   & $E_{2}$, mK & $a$, \AA  & $r_{0}$, \AA  & $E_{2}$, mK & $a$,\AA  & $r_{0}$, \AA\tabularnewline
\hline
HFD-B(He)   & 1.6853 (0.4\%)     & 88.60 (0.2\%)    & 7.28          & 1.6921      & 88.43    & 7.28 \tabularnewline
LM2M2       & 1.3034 (0.5\%)     & 100.2 (0.2\%)    & 7.33          & 1.3094      & 100.0    & 7.33 \tabularnewline
TTY         & 1.3149 (0.5\%)     & 99.82 (0.2\%)    & 7.32          & 1.3210      & 99.59    & 7.33 \tabularnewline
HFD-B3-FCII & 1.5872 (0.4\%)     & 91.19 (0.2\%)    & 7.29          & 1.5938      & 91.00    & 7.29 \tabularnewline
\hline
\end{tabular}
\caption{Binding energy and effective range parameters for different potential models calculated for exact and approximate effective coupling constants. Relative effect of rounding the coupling constant is given in parentheses.}
\label{tab:TwoBodyLowEnergyParameters}
\end{table}
Evidently, the binding energy and the scattering length are quite sensitive to the
often neglected small inaccuracy of the coupling constant.
Even though the the coupling constant employed in many published calculations
differs from the exact value only in the fifth figure, it is the third significant figure
of the binding energy and the scattering length which is affected by this small inaccuracy.

\begin{figure}
  \centerline{\includegraphics[width=0.8\textwidth,clip=true]{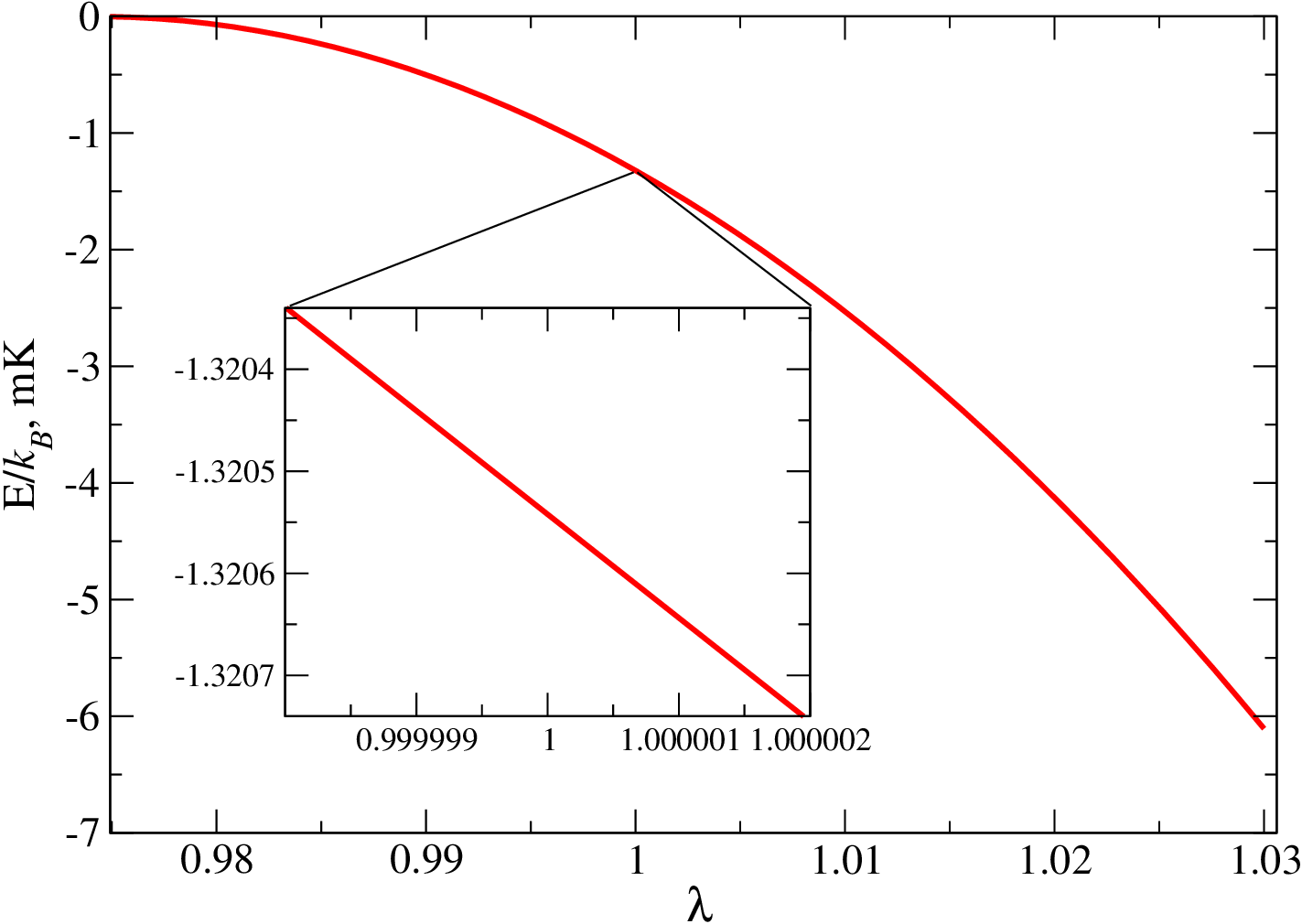}}
  \caption{He dimer binding energy as a function of the coupling constant.
          The inset shows the variation of the binding energy within the bounds determined by the inaccuracy of the fundamental constants. \label{fig:E2vsCC}}
\end{figure}

As the characteristics of near-threshold He-He states are quite sensitive to the accuracy of the effective coupling constant
employed, it is useful to place physically reasonable limits on the accuracy of few-body calculations.
Fig.~\ref{fig:E2vsCC} shows the coupling constant dependence of the He$_2$ binding energy dependence for the TTY potential.
The inset shows the variations of the two-body bound state energy within one-$\sigma$ confidence interval for the effective coupling
constant. This permits an estimation of the accuracy of the few-body near-threshold calculations at the level of 0.4~$\mu$K, which
corresponds to a relative error of the order of $\propto 3\times 10^{-4}$. This provides a natural limit to the accuracy of
numerical calculations for this system, a limit deriving from the known precision of the fundamental constants. Explicitly,
varying the effective coupling constant $\frac{\hbar^{2}}{m}$
from $12.11926$~K\AA$^{2}$ to $12.11930$~K\AA$^{2}$ we see the binding energy varying from
$-1.69231$~mK to $-1.69194$~mK. Thus, numerical calculations with
an accuracy of more than four significant figures are not possible, since the
potentials are defined in units of
the Boltzmann constant $k_{B}$.

\section{Three-body observables}
The known strong correlations between the two-body scattering length and the energies of trimer bound states suggests that we can also expect
sensitivity to the coupling constant in three-body calculations. The three-body results reported here were obtained using a general-purpose three-body code
based on solving Faddeev equations numerically. The numerical technique used in this code is similar to the one reported in
\cite{RoudnevFBS,RoudnevHeScat}. A very brief description of our approach to solving the Faddeev equations is given in the Appendix. The code is intended for public release and this work is a part of the release preparation.

\subsection{Three identical He atoms}
In order to determine the degree of sensitivity of three-body observables to small variations of the coupling constant,
we must first conduct a thorough analysis of the intrinsic numerical uncertainties and convergence properties when the
coupling constant is held fixed at its recommended value.
Convergence tables for bound states of the
\He{4}{3} trimer calculated with the LM2M2 potential are shown in Table~\ref{tab:LM2M2BoundConvergence}.
\begin{table}[!ht]
\begin{tabular}{lcccccc}
\hline
\multicolumn{7}{c}{Ground state energy, a.u.} \\
\hline
$N_x=N_y$ & $N_z=1$ & $N_z=2$ & $N_z=4$ & $N_z=6$ & $N_z=8$ & $\infty$ \\
\hline
20  &  -3.96752E-7 & -3.99142E-7 & -3.99137E-7  & -3.99342E-7 &  -3.99468E-7 & -3.99455e-07 \\
25  &  -3.98763E-7 & -3.99602E-7 & -4.00046E-7  & -4.00415E-7 &  -4.00356E-7 & -4.00360e-07 \\
30  &  -3.98560E-7 & -3.99610E-7 & -3.99967E-7  & -4.00139E-7 &  -4.00111E-7 & -4.00113e-07  \\
35  &  -3.98839E-7 & -4.00237E-7 & -4.00534E-7  & -4.00670E-7 &  -4.00671E-7 & -4.00671e-07 \\
40  &  -3.98781E-7 & -4.00235E-7 & -4.00476E-7  & -4.00547E-7 &  -4.00562E-7 & -4.00561e-07 \\
45  &  -3.98727E-7 & -4.00230E-7 & -4.00449E-7  & -4.00541E-7 &  -4.00565E-7 & -4.00563e-07 \\
50  &  -3.98764E-7 & -4.00310E-7 & -4.00513E-7  & -4.00577E-7 &  -4.00607E-7 & -4.00604e-07 \\
55  &  -3.98778E-7 & -4.00273E-7 & -4.00486E-7  & -4.00571E-7 &  -4.00595E-7 & -4.00593e-07 \\
60  &  -3.98753E-7 & -4.00298E-7 & -4.00494E-7  & -4.00567E-7 &  -4.00597E-7 & -4.00594e-07 \\
65  &  -3.98777E-7 & -4.00288E-7 & -4.00497E-7  & -4.00577E-7 &  -4.00603E-7 & -4.00601e-07 \\
70  &  -3.98763E-7 & -4.00298E-7 & -4.00496E-7  & -4.00572E-7 &  -4.00602E-7 & -4.00599e-07 \\
$\infty$
    &  -3.98767E-7 & -4.00296E-7 & -4.00498E-7  & -4.00574e-7 &  -4.00603E-7 & -4.00600e-07 \\
\hline
\multicolumn{7}{c}{Excited state energy, a.u.} \\
\hline
$N_x=N_y$ & $N_z=1$ & $N_z=2$ & $N_z=4$ & $N_z=6$ & $N_z=8$ & $\infty$ \\
\hline
20 & -7.20149E-9 & -7.21731E-9 & -7.21671E-9 & -7.21825E-9 &  -7.21908E-9 & -7.21911e-09 \\
25 & -7.20046E-9 & -7.20577E-9 & -7.20855E-9 & -7.21090E-9 &  -7.21046E-9 & -7.21047E-09 \\
30 & -7.20196E-9 & -7.20863E-9 & -7.21089E-9 & -7.21195E-9 &  -7.21172E-9 & -7.21172E-09 \\
35 & -7.20645E-9 & -7.21541E-9 & -7.21725E-9 & -7.21809E-9 &  -7.21806E-9 & -7.21806E-09 \\
40 & -7.20136E-9 & -7.21070E-9 & -7.21216E-9 & -7.21256E-9 &  -7.21262E-9 & -7.21263E-09 \\
45 & -7.20376E-9 & -7.21344E-9 & -7.21475E-9 & -7.21528E-9 &  -7.21541E-9 & -7.21541E-09 \\
50 & -7.20411E-9 & -7.21407E-9 & -7.21527E-9 & -7.21561E-9 &  -7.21578E-9 & -7.21578E-09 \\
55 & -7.20348E-9 & -7.21311E-9 & -7.21438E-9 & -7.21487E-9 &  -7.21499E-9 & -7.21500E-09 \\
60 & -7.20384E-9 & -7.21379E-9 & -7.21495E-9 & -7.21536E-9 &  -7.21555E-9 & -7.21555E-09 \\
65 & -7.20372E-9 & -7.21345E-9 & -7.21469E-9 & -7.21515E-9 &  -7.21531E-9 & -7.21531E-09 \\
70 & -7.20374E-9 & -7.21363E-9 & -7.21481E-9 & -7.21525E-9 &  -7.21543E-9 & -7.21543E-09 \\
$\infty$
   &  -7.20374E-9 & -7.21358E-9 & -7.21478E-9 & -7.21522E-9  & -7.21539E-9 & -7.21540e-09\\
\end{tabular}
  \caption{Convergence tables for \He{4}{3} bound state energies for LM2M2 potential.}
\label{tab:LM2M2BoundConvergence}
\end{table}
The bound state energies (with respect to the three-body break-up threshold)
are presented for different numbers of grid points in spline solutions to the Faddeev equations. The number of grid points in the ``cluster coordinate'' $x$
and in the ``reaction coordinate'' $y$ are set equal, and the number of grid points in the angular coordinate $z=({\bf x},{\bf y})$
is varied independently. (The detailed description of the coordinate system can be found in \cite{RoudnevFBS}). The cut-off distances are set to $R_x=1200$~a.u. for the cluster coordinate and $R_y=2000 \sqrt{\frac{3}{2}}$~a.u. for the reaction coordinate.
Not surprisingly, the excited state is much less sensitive to
the grid in angular coordinate: as is shown in \cite{RoudnevFBS}, the excited state is strongly dominated
by the two-body s-state.

To estimate the intrinsic numerical error we extrapolate the calculated values to
an infinite grid and compare the calculated values with the extrapolated one.
Estimating the error very conservatively we obtain the energy estimates
$E_{3}=(-4.0060\pm0.0001)\times 10^{-7}$~a.u.$=-126.499\pm0.003$~mK  for the ground state and
$E_{3}^{*}=(-7.21540\pm0.0001)\times 10^{-9}$~a.u.$=-2.2784\pm0.0003$~mK for the excited state with the LM2M2 potential.
Results obtained with other model potentials are given in Tab.~\ref{tab:He34BoundStates}.
\begin{table}[!ht]
\begin{tabular}{lcccccc}
\hline
Potential  & \multicolumn{3}{c}{Ground state energy} &  \multicolumn{3}{c}{Excited state energy} \\
           & \parbox{0.8in}{$E_3$, a.u. (best grid)} & \parbox{1in}{$E_3$, a.u. (extrapolated)} & \parbox{0.8in}{$E_3/k_B$, mK} &
             \parbox{0.8in}{$E^*_3$, a.u. (best grid)} & \parbox{1in}{$E^*_3$, a.u. (extrapolated)} & \parbox{0.8in}{$E^*_3/k_B$, mK} \\
\hline
HFD-B(He) & -4.21430E-7  & -4.21425E-7  & -133.075  & -8.68438E-9  & -8.68438E-9  & -2.74231  \\
LM2M2     & -4.00602E-7  & -4.00600E-7  & -126.499  & -7.21543E-9  & -7.21540E-9  & -2.27844  \\
TTY       & -4.00722E-7  & -4.00720E-7  & -126.537  & -7.25704E-9  & -7.25703E-9  & -2.29159 \\
HFDBFCI1  & -4.15369E-7  & -4.15365E-7  & -131.163  & -8.25861E-9  & -8.25861E-9  & -2.60786  \\
\hline
\end{tabular}
  \caption{Bound state energies (with respect to the break-up threshold) for the \He{4}{3} trimer.}
\label{tab:He34BoundStates}
\end{table}

We have also calculated the scattering length for a $^4$He atom scattered off the \He{4}{2} dimer. As our previous calculations have
shown \cite{RoudnevHeScat},
the box size is very critical for obtaining accurate results, and should be of order 3000~a.u. in the cluster coordinate. The results of
scattering length
calculations for the same set of potential models are shown in Tab.~\ref{tab:He34Scat}.
\begin{table}[!ht]
\begin{tabular}{lccc}
\hline
Potential  & \multicolumn{3}{c}{Scattering length estimates}  \\
           & \parbox{1.8in}{$a_{12}$, a.u. (best grid)} & \parbox{1.8in}{$a_{12}$, a.u. (extrapolated)} & \parbox{1.8in}{$a_{12}$, \AA } \\
\hline
HFD-B(He) & 230.416 & 230.424  & 121.93   \\
LM2M2     & 218.051 & 218.060  & 115.39   \\
TTY       & 218.936 & 218.945  & 115.86   \\
HFDBFCI1  & 228.380 & 228.388  & 120.86   \\
\hline
\end{tabular}
  \caption{Scattering length for the $^4{\rm He}$-\He{4}{2} atom-dimer scattering.}
\label{tab:He34Scat}
\end{table}

We can now address the issue of sensitivity to the coupling constant. How do small inaccuracies in the coupling constant used in many previous calculations affect the calculated observables?
For this purpose we have introduced a coupling constant $\lambda=12.11928/12.12=0.9999406$ in front of the potential and repeated the calculations using the most dense grid. In Tab~\ref{tab:He34Old} we compare results of such calculations with the results of previous calculations performed using similar techniques.
\begin{table}[!ht]
\begin{tabular}{lccc}
\hline
Potential  & \multicolumn{3}{c}{Scattering length estimates (rounded coupling constant)}  \\
           & \parbox{1.8in}{$a_{12}$, a.u. } & \parbox{1.8in}{$a_{12}$, \AA} & \parbox{1.8in}{$a_{12}$, \AA \cite{RoudnevHeScat}} \\
\hline
HFD-B(He) & 230.28 & 121.86 &  121.9  \\
LM2M2     & 217.74 & 115.22 &  115.4  \\
TTY       & 218.64 & 115.70 &  115.8  \\
HFDBFCI1  & 228.21 & 120.76 &   n/a \\
\hline
Potential  & \multicolumn{3}{c}{Ground state energy (rounded coupling constant)}  \\
           & \parbox{1.8in}{$E_3$, a.u. } & \parbox{1.8in}{$E_3/k_B$, mK} & \parbox{1.8in}{$E_3/k_B$, mK \cite{RoudnevFBS}} \\
\hline
HFD-B(He) & -4.21084E-7 & -132.968 &  -132.98  \\
LM2M2     & -4.00265E-7 & -126.394 &  -126.41  \\
TTY       & -4.00384E-7 & -126.431 &  -126.40  \\
HFDBFCI1  & -4.15689E-7 & -131.264 &  -131.26  \\
\hline
Potential  & \multicolumn{3}{c}{Excited state energy (rounded coupling constant)}  \\
           & \parbox{1.8in}{$E_3^*$, a.u. } & \parbox{1.8in}{$E_3^*/k_B$, mK} & \parbox{1.8in}{$E_3^*/k_B$, mK \cite{RoudnevFBS}} \\
\hline
HFD-B(He) & -8.65867E-9  & -2.7342 & -2.734   \\
LM2M2     & -7.19200E-9  & -2.2711 & -2.271  \\
TTY       & -7.23353E-9  & -2.2842 & -2.280   \\
HFDBFCI1  & -8.28580E-9  & -2.6164 & -2.617   \\
\hline
\end{tabular}
\caption{Calculated observables with the coupling constant modified to match previous calculations.}
  \label{tab:He34Old}
\end{table}
Examining Tab~\ref{tab:He34Old} we find that the results obtained here confirm the results of earlier calculations performed with
sparser meshes using a widely adopted inexact value of the coupling constant. The values obtained for the TTY potential, however,
differ from the previously reported values. As is mentioned above, in the previous calculations we used the value of the Boltzmann
constant suggested by the authors of the potential. Here we are using the value recommended by NIST.
It is also evident that the difference in three-body observables induced by the variation of the coupling constant at the level of
0.005\% is well resolved by our code. We have summarized the results of calculations performed with the exact and the rounded value of
the coupling constant in Table~\ref{tab:He3Summary}.
\begin{table}[!ht]
 \centering
 \begin{tabular}{ccccccc}
\hline
 & \multicolumn{3}{l}{$\frac{\hbar^{2}}{m}\equiv12.12\ $K\AA$^{2}$} & \multicolumn{3}{l}{$\frac{\hbar^{2}}{m}\equiv12.11928\
$K\AA$^{2}$}\tabularnewline
\hline
Potential   & $E_{3}$, mK &  $E_{3}^{*}$, mK   & $a_{12}$, \AA  &  $E_{3}$, mK  &  $E_{3}^{*}$, mK  & $a_{12}$, \AA\tabularnewline
\hline
HFD-B(He)   & -132.968 (0.08\%)     & -2.7342 (0.3\%)    & 121.86 (0.06\%)       & -133.075      & -2.74231    & 121.93 \tabularnewline
LM2M2       & -126.394 (0.08\%)     & -2.2711 (0.3\%)    & 115.22 (0.15\%)       & -126.499      & -2.27844    & 115.39 \tabularnewline
TTY         & -126.431 (0.08\%)     & -2.2842 (0.3\%)    & 115.70 (0.14\%)       & -126.537      & -2.29159    & 115.86 \tabularnewline
HFD-B3-FCII & -131.264 (0.08\%)     & -2.6164 (0.3\%)    & 120.76 (0.08\%)       & -131.163      & -2.60786    & 120.86 \tabularnewline
\hline
\end{tabular}
\caption{Energies of the bound states of \He{4}{3} and the atom-dimer scattering length for different potential models calculated for
exact and approximate effective coupling constants. In parethesis we show the relative difference between the results obtained with
exact and inexact coupling constant.}
 \label{tab:He3Summary}
\end{table}

\begin{figure}[htp]
  \includegraphics[width=0.8\textwidth, clip=true]{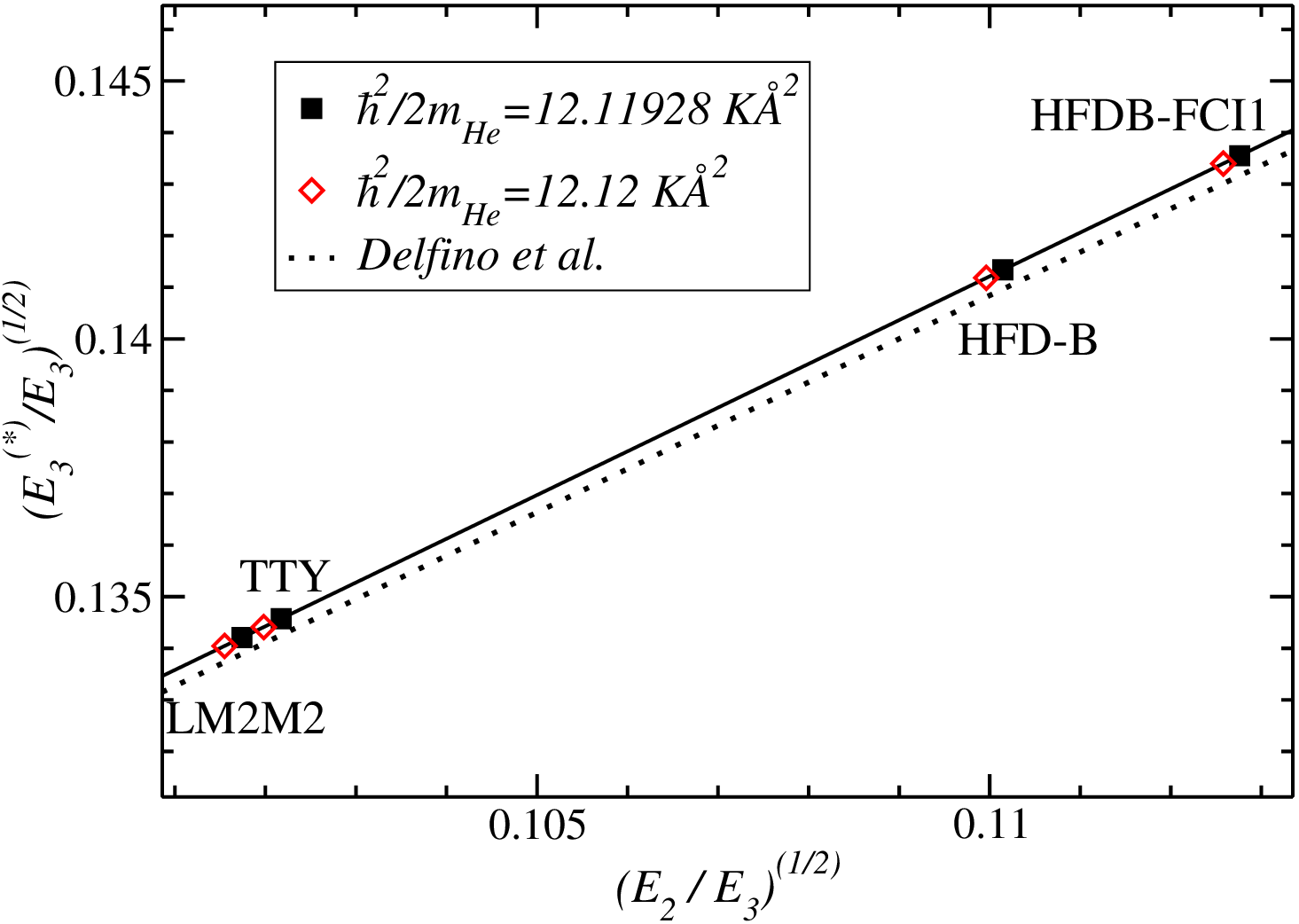}
\caption{Ratio of the energies of two subsequent bound states of the \He{4}{3} trimer
as a function of the dimer binding energy. The dotted line is recovered from Ref.~\cite{Delfino}
( with digitization error comparable with the observed discrepancy). The difference between the
results obtained with exact and inexact coupling constants is clearly resolved and consistent with the overall trend.}
\label{fig:DelfinoCorrelation}
\end{figure}

It has been suggested by many authors that the system of three He atoms should demonstrate a nearly universal behavior
\cite{Braaten,Penkov}.
Using a separable interaction Pen'kov has shown~\cite{Penkov}
that all the low-energy parameters of the system of three $^4$He atoms can be described
by a single dimensionless parameter within a few percent error. A similar observation has been made
by Braaten and Hammer~\cite{Braaten} on the basis of effective field theory and by
Platter and Phillips~\cite{PlatterPhillips} who used an approach similar to \cite{Braaten},
but used a higher order expansion and different regularization technique. In all these works
the pair angular momentum cut-off has been introduced and only the s-wave interaction has been taken into account.
Therefore, when comparing results of direct full-configuration calculations with predictions based on universality
arguments one can expect better agreement for the strongly s-wave dominated near-threshold states,
with poorer agreement expected for the ground state of the trimer -- which is not as dominated by the s-wave.
Each of these theories has a parameter which should be fit to reproduce
some three-body observable, and then predicts other three-body observables within a few percent error.
A similar observation was also made by Delfino {\it et al.}~\cite{Delfino},
who suggested that the ratio of neighboring bound state energies of an Efimov-like system
is a universal function of the ratio of the dimer binding energy to the energy of the three-body ground state.
We show this example in Fig.~\ref{fig:DelfinoCorrelation}. The curve from the original work~\cite{Delfino}
has been recovered graphically, and the accuracy of this procedure is comparable with the observed discrepancy.
It is, therefore, unclear, whether the small discrepancy comes from digitization or from the model employed in \cite{Delfino}.

\begin{figure}[htp]
  \centering
  \includegraphics[width=0.8\textwidth, clip=true]{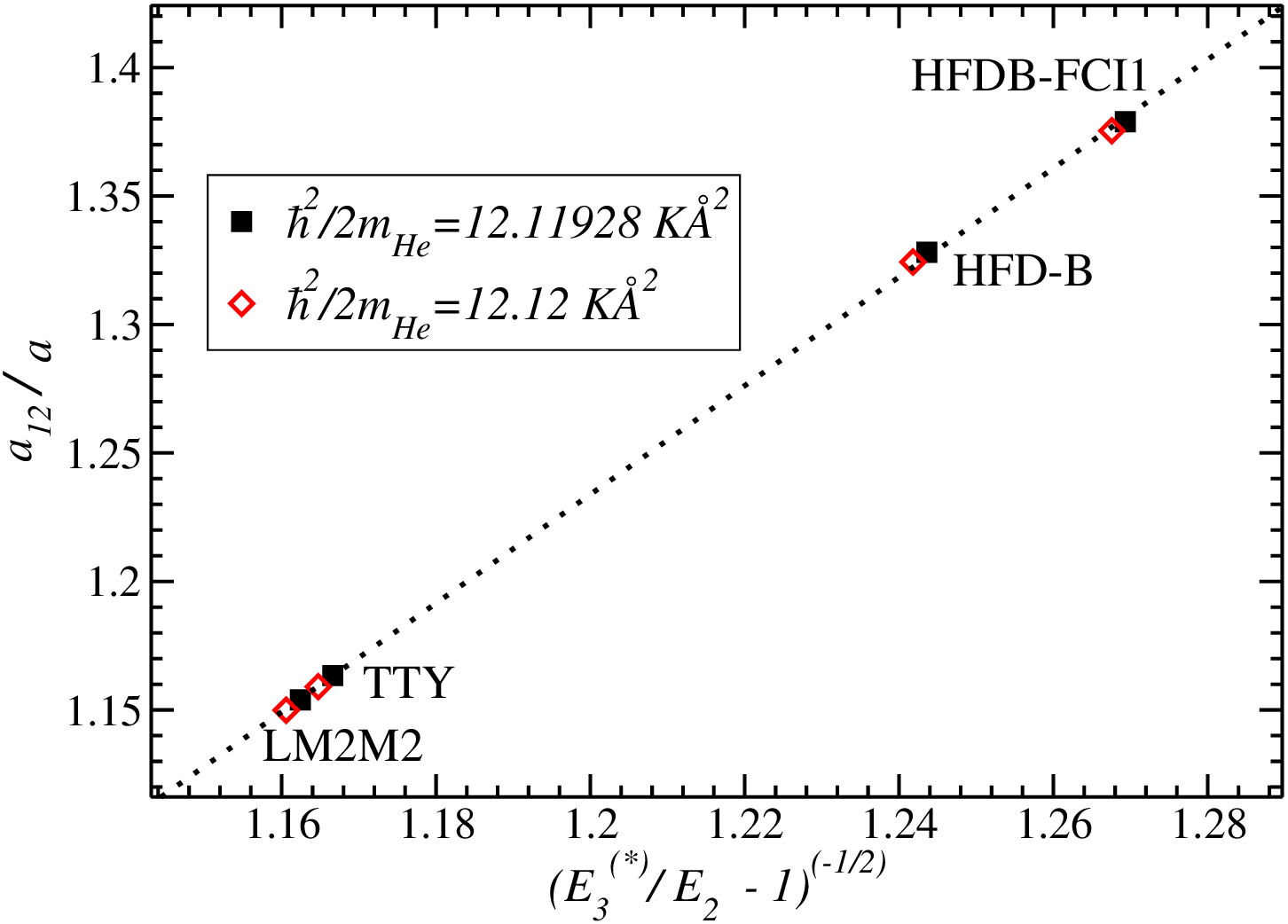}
  \caption{The atom-dimer scattering length (in units of the two-body scattering length) as a function of the distance to the pole in the three-body t-matrix. The difference between the results obtained with exact and inexact coupling constants is clearly resolved
  and consistent with the overall trend.}
  \label{fig:ScatDemonstration}
\end{figure}

Similar correlations can be observed for the atom-dimer scattering length. As the atom-dimer scattering is dominated by the
pole of the t-matrix corresponding to the near-threshold state of the trimer, the atom-dimer scattering length should behave as
\begin{equation}
  a_{12}\sim \frac{1}{\sqrt{E_2-E_3^*}} \sim \frac{a}{\sqrt{\frac{E_3^*}{E_2}-1}} ,
\label{eq:adScatLength}
\end{equation}
where $a$ is the two-body scattering length. Our numerical calculations confirm this simple observation extremely well.
In Fig.~\ref{fig:ScatDemonstration} we show the ratio of the atom-dimer scattering length
to the two-body scattering length as a function of the dimensionless parameter $\frac{1}{\sqrt{\frac{E_3^*}{E_2}-1}}$.
All the numerical results fall on a nearly perfect straight line.
(Similar connection between the neutron-deuteron doublet scattering length
and the energy of the three-body bound state is known in nuclear physics as the Phillips line \cite{PhillipsLine}).
Again, we see that the shifts in scattering length induced by an inaccuracy of the coupling constant consistently follow the overall trend.

Before turning to other examples of three-body systems that can be treated with our code, let us compare our results
with other known full-configuration calculations. We have summarized the results reported for the ground state of the
\He{4}{3} trimer with the TTY potential in Tab.~\ref{tab:He3Comparison}. In the left column we summarize the results obtained in
adiabatic hyperspherical (AH) representation, including single-channel \cite{ELG,VK} and multi-channel \cite{BG,NFJ,VK} calculations.
In the right column the results of direct numerical solution of the Schr\"odinger \cite{Yarevsky} or Faddeev equations
\cite{KolganovaMotovilov,Carbonell,RoudnevYakovlev} are given. The result reported in \cite{KolganovaMotovilov} was obtained
with a restricted angular basis corresponding to the very simplest angular grid used in the present calculations, and
agrees perfectly well with the value of -125.95 mK which we obtain in this simplified case.
The overall agreement between the results of solving the equations directly is much better than those obtained within the
AH approach. Although the AH approach provides an effective and reliable tool for studying
few-body systems qualitatively, obtaining converged results is more difficult. The variations of the
results due to inaccuracies in the conversion factors can not account for systematic underestimating of the binding energy
which we see in adiabatic hyperspherical calculations.
\begin{table}[ht]
  \centering
\begin{tabular}{clcl}
\hline
\multicolumn{2}{l}{ AH Reference} & \multicolumn{2}{l}{Direct calculation reference}         \\
\hline
\cite{ELG}    & -106.1 mK (a)      & \cite{KolganovaMotovilov}           & -125.9  mK \\
\cite{VK}     & -105 mK   (a)      & \cite{Carbonell}                    & -126.39 mK \\
\cite{BG}     & -125 mK   (b)      &  \cite{RoudnevYakovlev}             & -126.40 mK \\
\cite{NFJ}    & -125 mK   (b)      & \cite{Yarevsky}                     & -126.2 mK  \\
\cite{VK}     & -123.8$\pm$0.5 mK (b)  & Present                         & -126.537 (126.431) mK (c) \\
\hline
\end{tabular}
  \caption{Comparison of different results reported for \He{4}{3} ground state with TTY potential. (a) Single-channel
           approximation has been used. (b) Full multi-channel calculation in AH representation. (c) The value corresponding to the rounded coupling constant is given in parenthesis}
  \label{tab:He3Comparison}
\end{table}
We want to emphasize that although obtaining converged results in the AH approach is technically more difficult, it
should not be considered impossible. For instance, Suno and Esry \cite{SunoEsry2008} report
$E_3=-133.55$~mK for the ground state and $E_3^*=-2.7845$ mK for the excited state energy of the trimer in the calculation
with 35 AH channels using the SAPT2007 potential \cite{SAPT2007}. These values agree well with our result of $E_3=-133.589$~mK for the ground and $E_3^*=-2.78474$~mK for the excited state.

\section{Unsymmetric trimer}
Unlike the symmetric trimer, the trimer formed from two $^4$He atoms and one $^3$He atom has only one bound state.
In order to approach the Efimov regime there should be two subsystems possessing large binary scattering lengths compared to the effective range.
In the case of the unsymmetric trimer the $^3$He-$^4$He scattering length is only twice as big as the effective range,
and, therefore, the situation is quite far from the Efimov limit. The data published on the unsymmetric trimer is not as extensive as for the symmetric case.
However, the \He{4}{2}$^3$He trimer is a simple inhomogeneous three-body system, and as such it is an important
example for the benchmarking of three-body codes. In this section we report tests similar to those shown above for the symmetric system.

All the 2-body results are given in Table~\ref{tab:He3He4Binary}.
\begin{table}[!ht]
  \centering
\begin{tabular}{ccccc}
\hline
 & \multicolumn{2}{l}{$\frac{\hbar^{2}}{m}\equiv12.12\ $K\AA$^{2}$} & \multicolumn{2}{l}{$\frac{\hbar^{2}}{m}\equiv12.11928\
$K\AA$^{2}$}\tabularnewline
\hline
Potential   & $a$, a.u.  & $r_{0}$, a.u. & $a$, a.u.  & $r_{0}$, a.u.\tabularnewline
\hline
HFD-B(He)   & -34.382 & 18.40    & -34.401 & 18.39 \tabularnewline
LM2M2       & -33.245 & 18.56    & -33.263 & 18.56 \tabularnewline
TTY         & -33.226 & 18.56    & -33.244 & 18.55 \tabularnewline
HFD-B3-FCII & -34.050 & 18.45    & -34.069 & 18.45 \tabularnewline
\hline
\end{tabular}
  \caption{Scattering length and the effective range for $^3$He-$^4$He collisions.}
  \label{tab:He3He4Binary}
\end{table}
As we can see from Table~\ref{tab:He3He4Binary}, the $^3$He-$^4$He scattering length is less sensitive
to the variations of the coupling constant, so the variations in three-body observables should mostly be due to
the sensitivity of the \He{4}{2} subsystem. The three-body data is summarized in Table~\ref{tab:He3He4Summary}.
The sensitivity of the unsymmetric trimer bound state energy to small variations of the coupling constant is comparable to the sensitivity of the bosonic trimer excited state. The atom-dimer scattering length, however, is much less sensitive.
\begin{table}[!ht]
  \begin{tabular}{ccccc}
    \hline
       & \multicolumn{2}{l}{$\frac{\hbar^{2}}{m}\equiv12.12\ $K\AA$^{2}$}
                     & \multicolumn{2}{l}{$\frac{\hbar^{2}}{m}\equiv12.11928\ $K\AA$^{2}$} \tabularnewline
    \hline
    Potential   & $E_3$, a.u.  & $a_{12}$, a.u. & $E_3$, a.u.  & $a_{12}$, a.u.\tabularnewline
    HFD-B(He)   & -5.3815E-8 (0.3\%) & 36.18 (0.03\%) & -5.3958E-8 & 36.17 \tabularnewline
    LM2M2       & -4.5355E-8 (0.3\%) & 36.91 (0.03\%) & -4.5488E-8 & 36.90 \tabularnewline
    TTY         & -4.5270E-8 (0.3\%) & 37.23 (0.03\%) & -4.5404E-8 & 37.24 \tabularnewline
    HFD-B3-FCII & -5.1442E-8 (0.3\%) & 36.57 (0.03\%) & -5.1582E-8 & 36.56 \tabularnewline
    \hline
  \end{tabular}
  \caption{\He{4}{2}$^3$He bound state energy and $^3$He-\He{4}{2} atom-dimer scattering length. In parenthesis we show the relative difference between the results obtained with exact and inexact coupling constant.}
  \label{tab:He3He4Summary}
\end{table}

\begin{figure}[ht]
 \centering
 \includegraphics[width=0.8\textwidth,clip=true]{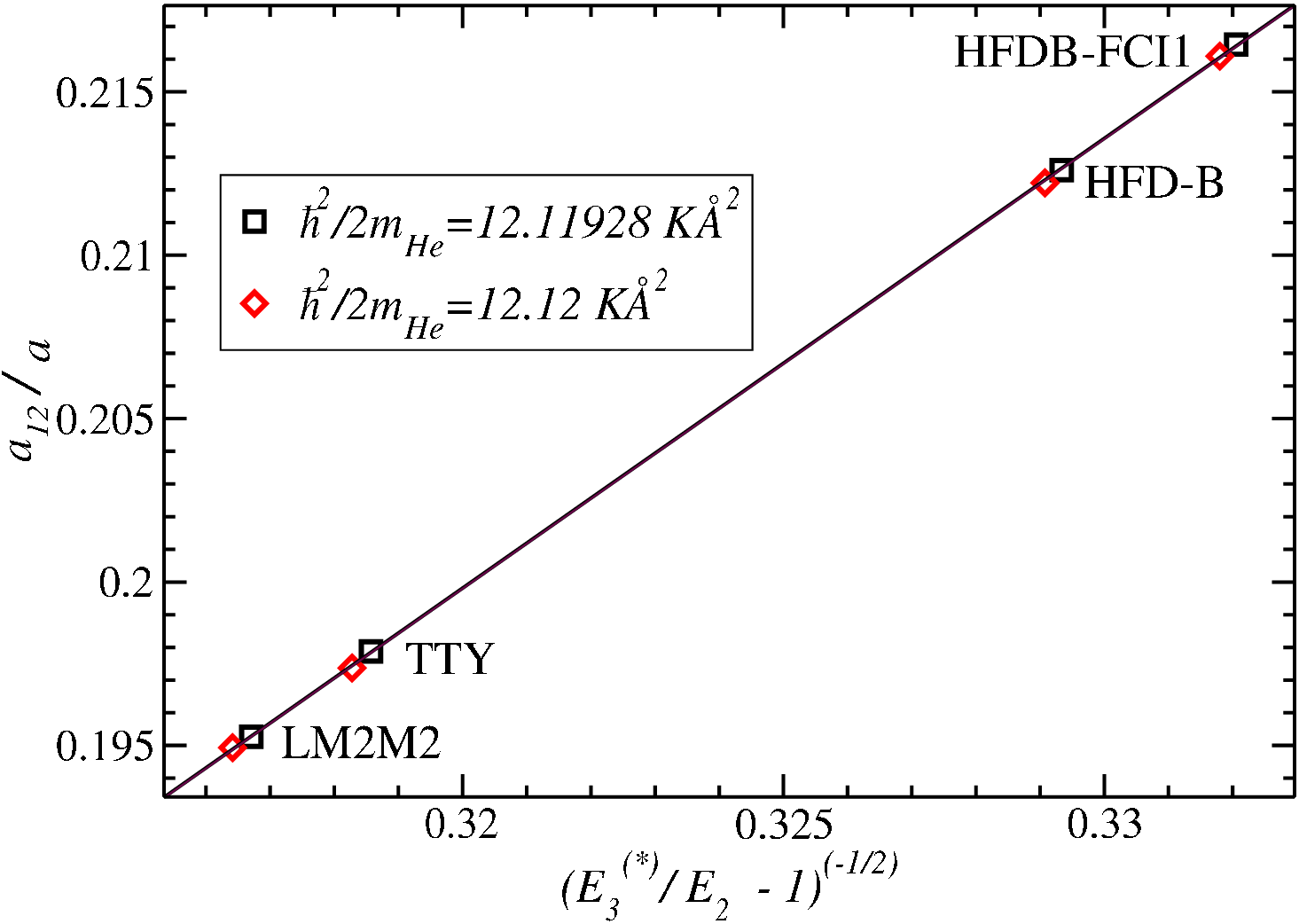}
 \caption{The $^3$He-\He{4}{2} atom-dimer scattering length (in units of the $^4$He-$^4$He scattering length) as a function of the distance to the pole in the three-body t-matrix. The difference between the results obtained with exact and inexact coupling constants is clearly resolved and consistent with the overall trend.}
 \label{fig:UnsymmTrend}
\end{figure}

Similarly to the case of the symmetric bosonic trimer, we have plotted the ratio of the $^3$He-\He{4}{2} atom-dimer scattering length to the $^4$He-$^4$He scattering length as a function of the dimensionless parameter $\frac{1}{\sqrt{\frac{E_3}{E2}-1}}$ which characterizes the distance between the two-body threshold and the pole of the three-body t-matrix corresponding to the unsymmetric trimer bound state. All the data consistently follow the same trend (see Fig.~\ref{fig:UnsymmTrend}).

\section{Conclusions}
We have performed highly accurate calculations of the system of three He atoms, including both symmetric and unsymmetric cases.
We have studied the effects of small (0.006\%) variations of the interaction coupling constant on the energies of He dimer and trimer
bound states and scattering lengths. Results of the calculations reported here are converged to 5 significant figures for the most of
the observables. All the results are consistent with previously published calculations and follow the same universal -- potential
independent -- trend.
We can summarize the sensitivity of the results to small inaccuracies in the coupling constant as follows:
a 0.006\% variation of the coupling constant induces
0.5\% shift in the energy of the dimer,
0.3\% shift in the energy of the excited state of the homogeneous trimer and the single bound state of the inhomogeneous trimer,
0.2\% shift in the $^4$He-$^4$He atom-atom scattering length,
0.06\%-0.15\% shift in the $^4$He-\He{4}{2} atom-dimer scattering length,
0.08\% shift in the energy of the ground state of the homogeneous trimer,
and 0.03\% shift in the scattering length for $^3$He-\He{4}{2} collisions.
As the effective coupling constant for the potentials reported in units of temperature is known with the relative
accuracy of 0.0002\%, the physical limit for the accuracy of the binding energy of the He dimer, the excited state of homogeneous He trimer and the energy of the inhomogeneous He trimer is about 5 significant figures. This accuracy is achieved in the present calculations performed with a computer program being prepared for a public release.

The authors hope that both the qualitative observations and the numerical results reported in this work can be used in the future for benchmarking various quantum few-body codes.

\section*{Acknowledgements}
This work is supported by the NSF grant PHY-0903956. We wish to thank Dr. Kolganova (JINR, Dubna) for
stimulating discussions and independent preliminary testing of the three-body code.

\section*{Appendix A}
Here we provide a very brief overview of our approach to
solving the Faddeev equations numerically. More detailed and rigorous
description of the equations and the numerical approach is being prepared
as a separate publication.

According to the Faddeev formalism \cite{FaddMerk} the wave function
of three particles is expressed in terms of Faddeev components $\Phi$
\[
\Psi(\mathbf{x_{1}},\mathbf{y_{1}})=\Phi_{1}(\mathbf{x_{1}},\mathbf{y_{1}})+\Phi_{2}(\mathbf{x_{2}},\mathbf{y_{2}})+\Phi_{3}(\mathbf{x_{3}},\mathbf{y_{3}})\:,
\]
where $\mathbf{x}_{\alpha}$ and $\mathbf{y}_{\alpha}$ are Jacobi
coordinates corresponding to the fixed pair $\alpha$
\begin{equation}
\begin{array}{c}
  \mathbf{x}_{\alpha}=(\frac{2m_{\beta}m_{\gamma}}{m_{\beta}+m_{\gamma}})^{\frac{1}{2}}(\mathbf{r}_{\beta}-\mathbf{r}_{\gamma})\ ,\\
  \mathbf{y}_{\alpha}=(\frac{2m_{\alpha}(m_{\beta}+m_{\gamma})}{m_{\alpha}+m_{\beta}+m_{\gamma}})^{\frac{1}{2}}(\mathbf{r}_{\alpha}-\frac{m_{\beta}\mathbf{r}_{\beta}+m_{\gamma}\mathbf{r}_{\gamma}}{m_{\beta}+m_{\gamma}})\ .\end{array}
\label{eq:Jacoord}
\end{equation}
Here $\mathbf{r}_{\alpha}$ are the positions of the particles in
the center-of-mass frame. The Faddeev components obey the set of three
equations
\begin{equation}
\begin{array}{c}
\displaystyle (H_{0}+V_{\alpha}(\mathbf{x}_{\alpha})-E)\Phi_{\alpha}(\mathbf{x}_{\alpha},\mathbf{y}_{\alpha})=
                       -V_{\alpha}(x_{\alpha})\sum_{\beta\neq\alpha}\Phi_{\beta}(\mathbf{x}_{\beta},\mathbf{y}_{\beta})\\
\alpha=1,2,3\end{array}\,,
\label{eqf}
\end{equation}
where $V_{\alpha}(\mathbf{x}_{\alpha})$ stands for the pairwise potential
and $H_{0}$ is the kinetic energy of the three particles. To make
this system of equations suitable for numerical calculations one should
take into account the symmetries of the physical system. As far as
all the model potentials are central it is possible to factor out
the degrees of freedom corresponding to the rotations of the whole
cluster \cite{TAM}. For the case of zero total angular momentum the
reduced Faddeev equation reads
\begin{equation}
{\displaystyle (H_{0\alpha}+V_{\alpha}(x_{\alpha})-E)\Phi_{\alpha}(x_{\alpha},y_{\alpha},z_{\alpha})=-x_{\alpha}y_{\alpha}V_{\alpha}(x_{\alpha})\sum_{\beta\neq\alpha}\frac{1}{x_{\beta}y_{\beta}}\Phi_{\beta}(x_{\beta},y_{\beta},z_{\beta})\:.}
\label{Fadd3}
\end{equation}
Here
$H_{0\alpha}=-\frac{\partial^{2}}{\partial x_{\alpha}^{2}}-\frac{\partial^{2}}{\partial y_{\alpha}^{2}}-(\frac{1}{x_{\alpha}^{2}}+\frac{1}{y_{\alpha}^{2}})\frac{\partial}{\partial z_{\alpha}}(1-z_{\alpha}^{2})^{\frac{1}{2}}\frac{\partial}{\partial z_{\alpha}}$,
\begin{equation}
\begin{array}{c}
x_{\alpha}=|\mathbf{x}_{\alpha}|\,,\\
y_{\alpha}=|\mathbf{y_{\alpha}}|\,,\\
{\displaystyle z_{\alpha}=\frac{(\mathbf{x_{\alpha}},\mathbf{y_{\alpha}})}{x_{\alpha}y_{\alpha}}\,,}\end{array}
\label{intrcoord}
\end{equation}
and the coordinate transformations between different system of Jacobi
coordinates follow from the definition of the Jacobi coordinates (\ref{eq:Jacoord}).

The asymptotic boundary condition for bound states consists of two
terms \cite{FaddMerk}
\[
\Phi(x,y,z)\sim\:\phi_{2}(x)e^{-k_{y}y}+A(\frac{x}{y},z)\frac{e^{-k_{3}(x^{2}+y^{2})^{\frac{1}{2}}}}{(x^{2}+y^{2})^{\frac{1}{4}}}\:,
\]
where $\phi_{2}(x)$ is the two-body bound state wave function, $k_{y}=\sqrt{E_{2}-E_{3}}$,
$k_{3}=\sqrt{-E_{3}}$, $E_{2}$ is the energy of the two-body bound
state and $E_{3}$ is the energy of the three-body system. The second
term -- which corresponds to virtual decay of the three body bound state into
three free particles -- decreases much faster than the first one which
corresponds to virtual decay into a particle and a two-body cluster.
In our calculations we neglect the second term in the asymptotic
introducing the following approximate boundary conditions for the
Faddeev component at sufficiently large distances $R_{x}$ and $R_{y}$
\begin{equation}
\begin{array}{l}
{\displaystyle \frac{\partial_{x}\Phi(x,y,z)\lfloor_{x=R_{x}}}{\Phi(x,y,z)\lfloor_{x=R_{x}}}=k_{2}\equiv i\sqrt{E_{2}}\:,}\\
{\displaystyle \frac{\partial_{_{y}}\Phi(x,y,z)\lfloor_{y=R_{y}}}{\Phi(x,y,z)\lfloor_{y=R_{y}}}=k_{y}\:.}\end{array}
\label{bc}
\end{equation}

In order to solve the equations numerically we introduce a basis of
Hermit splines satisfying the boundary conditions (\ref{bc}) and
use orthogonal collocations to calculate a discrete matrix analog
of the Faddeev operator
\[
(\hat{H}_{0\alpha}+\hat{V}_{\alpha}\hat{S}_{\alpha}-\hat{S}_{\alpha}E)\hat{\Phi}_{\alpha}=-\hat{V}_{\alpha}(\hat{C}_{\alpha\beta}\hat{\Phi}_{\beta}+\hat{C}_{\alpha\gamma}\hat{\Phi}_{\gamma})\ \ .
\]
More detailed description of the discretization procedure can be found in \cite{RoudnevFBS} (see also \cite{GridGen} where
we describe the procedure of constructing optimal non-uniform grids automatically).

We solve the system of linear equations iteratively exploiting factorability
of the left-hand side of Eqs. (\ref{Fadd3}) for preconditioning.
In particular, we introduce localized components
\begin{equation}
  \tau_{\alpha}\equiv(\hat{H}_{0\alpha}+\hat{V}_{\alpha}\hat{S}_{\alpha}-\hat{S}_{\alpha}E)\hat{\Phi}_{\alpha}
  \label{eq:Tau}
\end{equation}
which -- due to the asymtotic properties of the Faddeev components
-- has much better spacial localization than the original Faddeev
component. The equations for the localized component read
\begin{equation}
\tau_{\alpha}=-\hat{V}_{\alpha}(\hat{C}_{\alpha\beta}(\hat{H}_{0\beta}+\hat{V}_{\beta}\hat{S}_{\beta}-\hat{S}_{\beta}E)^{-1}\tau_{\beta}+\hat{C}_{\alpha\gamma}(\hat{H}_{0\gamma}+\hat{V}_{\gamma}\hat{S}_{\gamma}-\hat{S}_{\gamma}E)^{-1}\tau_{\gamma})
\label{eq:LCFadd}
\end{equation}
The transformation (\ref{eq:Tau}) makes it possible to reduce the rank of the linear
system essentially, typically by a factor of 3. Application of Eqs. (\ref{eq:LCFadd}) to atom-dimer scattering can be found in \cite{RoudnevHeScat}. A similar idea has been discussed in \cite{Rawitscher}.


\begin{thebibliography}{00}
\bibitem{HeExp1} F. Luo, G. C. McBane, G. Kim, C. F. Giese, and W. R.
Gentry, J. Chem. Phys. {\bf 98}, 3564 (1993).
\bibitem{HeExp2} W. Schöllkopf and J. P. Toennies, Science {\bf 266}, 1345
(1994).
\bibitem{HeExp3} R. E. Grisenti {\em et al.}, Phys. Rev. Lett. {\bf 85}, 2284 (2000).
%
\bibitem{Az79} R. A. Aziz, V. P. S. Nain, J. S. Carley, W. L. Taylor, and G. T. McConville, J. Chem. Phys. {\bf 70}, 4330 (1979).
\bibitem{HFD-B} R. A. Aziz, F. R. W. McCourt and C. C. K. Wong,  Mol. Phys. {\bf 61}, 1487 (1987).
\bibitem{LiuMcLean}B. Liu and A. D. McLean, J. Chem. Phys. {\bf 91}, 2348 (1989).
\bibitem{LM2M2} Ronald A. Aziz and Martin J. Slaman, J. Chem Phys, {\bf 94}, 8047 (1991).
\bibitem{TTY} K.T.Tang, J.P.Toennies and C.L.Yiu, Phys. Rev. Lett. {\bf 74}, 1546 (1995).
\bibitem{VanMourik} T. van Mourik and J. H. van Lenthe, J. Chem. Phys. {\bf 102}, 7479 (1995).
\bibitem{SAPT2007} M. Jeziorska, W. Cencek, B. Patkowski, B. Jeziorski, and K. Szalewicz, J. Chem. Phys. {\bf 127}, 124303 (2007).
%
\bibitem{Pandharipande} V. R. Pandharipande, J. G. Zabolitzky, S. C.
Pieper, R. B. Wiringa, and U. Helmbrecht, Phys. Rev. Lett. {\bf 50}, 1676 (1983).
\bibitem{Glockle} Th. Cornelius and W. Gl\"ockle, J. Chem. Phys, {\bf 85}, 3906 (1986).
\bibitem{ELG} B. D. Esry, C. D. Lin, and C. H. Greene, Phys. Rev. A \textbf{54}, 394 (1996).
\bibitem{NFJ} E. Nielsen, D. V. Fedorov, and A. S. Jensen, J. Phys. B \textbf{31}, 4085 (1998).
\bibitem{MSK97} A. K. Motovilov, S. A. Soﬁanos, and E. A. Kolganova, Chem. Phys. Lett. {\bf 275}, 168 (1997).
\bibitem{KMS98} E. A. Kolganova, A. K. Motovilov, and S. A. Soﬁanos, J. Phys. B {\bf 31}, 1279–1302 (1998).
\bibitem{KM99}  E. A. Kolganova and A. K. Motovilov, Phys. At. Nucl. {\bf 62}, 1179 (1999).
 \bibitem{RoudnevYakovlev} V. Roudnev and S. Yakovlev, Chem. Phys. Lett. \textbf{328}, 97–106 (2000)
\bibitem{BG} D. Blume and C. H. Greene, J. Chem. Phys. \textbf{113}, 2145 (2000)
\bibitem{HeNeAr} D. Blume, Chris H. Greene, B.D. Esry, J. Chem.Phys., \textbf{6}, 2145 (2000)
\bibitem{Filikhin} I.N.Filikhin, S.L. Yakovlev, V.A. Roudnev and B.Vlahovic, J. Phys. B {\bf 35}, 501 (2001).
\bibitem{RoudnevHeScat} V. Roudnev, Chem. Phys. Lett. \textbf{367}, 95–101 (2003)
\bibitem{Braaten} E. Braaten and H.-W. Hammer, Phys. Rev. A, {\bf 67}, 042706 (2003)
\bibitem{Penkov} F. M. Pen'kov, Journal of Experimental and Theoretical Physics, {\bf 97}, 485 (2003)
\bibitem{RoudnevFBS} V. A. Roudnev, S. L. Yakovlev, and S. A. Sofianos, Few-Body Systems \textbf{37}, 179–196 (2005)
\bibitem{Carbonell} Rimantas Lazauskas and Jaume Carbonell, Phys.Rev. A \textbf{73}, 062717 (2006)
\bibitem{PlatterPhillips} L. Platter and D.R. Phillips, Few-Body Systems, {\bf 40}, 35 (2006)
\bibitem{VK} Viatcheslav Kokoouline and Françoise Masnou-Seeuws, Phys. Rev. A \textbf{73}, 012702 (2006)
\bibitem{Yarevsky} Moses Salci, Evgeny Yarevsky, Sergey B. Levin, Nils Elander, International Journal of Quantum Chemistry, \textbf{107}, 464–468 (2007)
\bibitem{SunoEsry2008} Hiroya Suno and B. D. Esry, Phys. Rev. A 78, 062701 (2008)
\bibitem{KolganovaMotovilov} E. A. Kolganova, A. K. Motovilov, and W. Sandhas, Physics of Particles and Nuclei {\bf 40}, 206 (2009).
%
\bibitem{Delfino} A. Delfino, T. Frederico, and L. Tomio, Few-Body Systems, {\bf 28}, 259 (2000)
\bibitem{PhillipsLine} A.C. Phillips Nucl. Phys A {\bf 107}, 209 (1968)
\bibitem{FaddMerk} L.D. Faddeev, S.P. Merkuriev, \emph{Quantum scattering
theory for several particle systems} (Doderecht: Kluwer Academic Publishers, (1993))
\bibitem{TAM} V.V. Kostrykin,~A.A.~Kvitsinsky,S.P.~Merkuriev, Few-Body~Systems, \textbf{6}, 97, (1989)
\bibitem{GridGen} V. Roudnev and Michael Cavagnero, Comp. Phys. Comm., \textbf{182} 2099–2106 (2011).
\bibitem{Rawitscher} W. Gloeckle and G. Rawitscher, arXiv:physics/0512010 (2005)
\end{thebibliography}
\end{document}